\documentclass[aps,prd,preprint]{revtex4-1}
\usepackage {graphicx}
\usepackage[toc,page]{appendix}
\usepackage{amsmath}
\newcommand{\be}{\begin{equation}}
\newcommand{\ee}{\end{equation}}
\begin{document}
\title{Cosmological initial data for numerical relativity}
\author{David Garfinkle}
\email{garfinkl@oakland.edu}
\affiliation{Dept. of Physics, Oakland University, Rochester, MI 48309, USA}
\affiliation{Leinweber Center for Theoretical Physics, Randall Laboratory of Physics, University of Michigan, Ann Arbor, MI 48109-1120, USA}
\author{Lawrence Mead}
\email{Lawrence.Mead@usm.edu}
\affiliation{Dept. of Physics and Astronomy, University of Southern Mississippi, Hattiesburg, MS 39406, USA}

\date{\today}

\begin{abstract}
We find initial data for numerical relativity simulations of inhomogeneous cosmologies.  This involves treating an exceptional case of the general relativity constraint equations.  We devise analytic and numerical methods to treat this exceptional case.  We apply the analytic method to the standard case of cosmology with a single scalar field. The numerical method is applied to the two-field ekpyrotic cosmology.

\end{abstract}


\maketitle

\section{Introduction}

Numerical relativity simulations of inhomogeneous cosmologies are performed for a variety of reasons: to explore the inflationary scenario\cite{east1,clough1,clough2,lehner}, or the ekpyrotic scenario\cite{ekpsim1,ekpsim2,anna1,anna2}, or the nature of spacetime singularities\cite{beverly1,beverly2,allofus,dgharmonic,dgprl}, or cosmological structure formation.\cite{east2,starkman,durrer}  Any simulation must start with initial data, which in general relativity entails solving coupled nonlinear constraint equations.\cite{York} 

This is very different from the usual treatments of inhomogeneous cosmologies.  There the inhomogeneities are typically treated in first order perturbation theory.  This allows the perturbations to be separated into modes that decouple and thus can each be treated independently.  The initial data can essentially be specified freely.

We would like to have numerical relativity initial data of sufficient generality that it essentially corresponds to the sort of initial data used in cosmological perturbation theory.  This leads to difficulties, since that sort of data corresponds to an exceptional case in the treatment of the relativity constraint equations.  However, we present a method to overcome these difficulties.

In section II we present the constraint equations of general relativity.  In section III we specialize to the case relevant to cosmology and show how to overcome the difficulties associated with this exceptional case.

Section IV presents the application of our method to finding numerical relativity initial data that are as close as possible to standard one-field cosmological perturbations.  Section V presents a more challenging case associated with the two-field ekpyrotic scenario.  Our conclusions are given in section VI.  

\section{Constraint Equations}

Initial data for a numerical relativity simulation consists of a three dimensional manifold $\Sigma$ on which
there is a spatial metric $\gamma_{ij}$ and an extrinsic curvature $K_{ij}$.  Here $\Sigma$ represents all of space at the initial time at which the simulation starts.  In a phase space picture, 
$\gamma_{ij}$ is the configuration variable and $K_{ij}$ is the momentum variable.  The data cannot be freely specified, but instead must satisfy two equations called the momentum constraint
\begin{equation}
{D^i}{K_{ij}} - {D_j}K = -  {{\gamma^i}_j}{T_{i\mu}}{n^\mu}
\label{mom1}
\end{equation}
and the Hamiltonian constraint
\begin{equation}
{^{(3)}}R + {K^2} - {K^{ij}}{K_{ij}} = 2 {T_{\mu \nu}}{n^\mu}{n^\nu}
\label{ham1}
\end{equation}
Here $n^\mu$ is the normal to the initial data surface, $D_i$ is the spatial covariant derivative, and $^{(3)}R$ is the spatial scalar curvature. $T_{\mu \nu}$ is the stress-energy tensor and we have chosen units where $8\pi G = 1$.  Initial data must also be specified for the matter fields that make up $T_{\mu \nu}$. 

It is helpful to decompose the extrinsic curvature into its trace $K$ and a trace-free part $A_{ij}$ given by 
\begin{equation}
{A_{ij}} = {K_{ij}} - {\textstyle {\frac 1 3}} K {\gamma _{ij}}
\end{equation}
Then the constraint equations become
\begin{eqnarray}
{D^i}{A_{ij}} - {\textstyle {\frac 2 3}} {D_j}K = -  {{\gamma^i}_j}{T_{i\mu}}{n^\mu}
\label{mom2}
\\
{^{(3)}}R + {\textstyle {\frac 2 3}} {K^2} - {A^{ij}}{A_{ij}} = 2 {T_{\mu \nu}}{n^\mu}{n^\nu}
\label{ham2}
\end{eqnarray}

The constraint equations are usually solved by the York method.\cite{York}  This method begins by introducing rescaled quantities  
${\tilde \gamma} _{ij}$ and ${\tilde A}_{ij}$ given by
\begin{equation}
{{\tilde \gamma} _{ij}} = {\psi ^{-4}} {\gamma_{ij}}
\label{gammatilde}
\end{equation}
and ${{\tilde A}_{ij}} = {\psi ^2}{A_{ij}}$.  
The quantity ${\tilde A}_{ij}$ is then expressed as 
\begin{equation}
{{\tilde A}_{ij}} =  {X_{ij}} + {{\tilde D} _i}{W_j}+{{\tilde D} _j}{W_i} - {\textstyle {\frac 2 3}}
{{\tilde \gamma} _{ij}} {{\tilde \gamma} ^{mn}}{{\tilde D}_m}{W_n} 
\label{Atilde}
\end{equation}
It seems odd to introduce these new quantities $\psi$ and $W_i$.  However, as we will soon see, they are essentially ``correction terms'' to be used to convert an initial guess for a solution of the constraint equations into an actual solution.

Using eqns.(\ref{gammatilde}-\ref{Atilde}) in eqns.(\ref{mom2}-\ref{ham2}) we obtain
\begin{eqnarray}
{{\tilde D}^i} \left ({{\tilde D} _i}{W_j}+{{\tilde D} _j}{W_i} - {\textstyle {\frac 2 3}}
{{\tilde \gamma} _{ij}} {{\tilde D}^k}{W_k} \right ) + {{\tilde D}^i}{X_{ij}} 
- {\textstyle {\frac 2 3}} {\psi^6}{D_j}K = - {\psi^6} {{\gamma^i}_j}{T_{i\mu}}{n^\mu}
\label{mom3}
\\
{{\tilde D}^i}{{\tilde D}_i} \psi \; - \; {\frac 1 8} ({^{(3)}}{\tilde R})\psi \; - \; {\frac 1 {12}}
{K^2} {\psi ^5} \; + \; {\frac 1 8} {{\tilde A}^{ij}}{{\tilde A}_{ij}} {\psi ^{-7}} = - \; {\frac 1 4}  {T_{\mu \nu}}{n^\mu}{n^\nu} {\psi ^5}
\label{ham3}
\end{eqnarray}
Here spatial indices are raised and lowered with ${\tilde \gamma}_{ij}$.  The derivative operator ${\tilde D}_i$ and scalar ${^{(3)}}{\tilde R}$ are respectively the covariant derivative and scalar curvature associated with ${\tilde \gamma}_{ij}$.

For our purposes, it is helpful to think of the quantities used in the York method as follows: $K$ is to be freely specified.  ${\tilde \gamma}_{ij}$ and $X_{ij}$ are our initial guesses for 
$\gamma _{ij}$ and $A_{ij}$.  That is, if we happened to have $({\gamma_{ij}},{A_{ij}})$ satisfying eqns.(\ref{mom2}-\ref{ham2}) then the choice $\psi=1$ and ${W_i}=0$ would solve eqns. (\ref{mom3}-\ref{ham3}).  If our initial guess does not solve the constraint equations, then $W_i$ and $\psi$ are correction terms that turn our initial guess into a solution. That is, by solving eqns. (\ref{mom3}-\ref{ham3}) for $W_i$ and $\psi$ we obtain a solution of eqns. (\ref{mom2}-\ref{ham2}).  So our task of solving the constraint equations has reduced to the task of solving eqns. (\ref{mom3}-\ref{ham3}) for $W_i$ and $\psi$.

As it stands, eqns. (\ref{mom3}-\ref{ham3}) are coupled, nonlinear differential equations.  However, the standard procedure decouples them as follows: first define the quantity ${\tilde J}_j$ by
\begin{equation}
{{\tilde J}_j} = {\psi^6} {{\gamma^i}_j}{T_{i\mu}}{n^\mu}
\label{current}
\end{equation} 
For each choice of matter fields, we must choose a way of specifying initial data so that ${\tilde J}_j$ does not depend on $\psi$.  In section IV we will give an explicit example of how to perform this sort of specification.  

Second, choose $K$ to be constant, so that ${{\tilde D}_i}K=0$.  
This choice of $K$ to be constant sounds like a loss of generality in the choice of initial data, but it turns out that it's not, for the following reason: the result of evolving the initial data in a numerical  relativity simulation will be a spacetime.  But spacetime can be divided up into space and time in many different ways.  One such way is to have the surfaces of constant time be surfaces of constant $K$.  So in choosing constant $K$ for our initial data surface, we are simply making use of the coordinate invariance of general relativity.  Or to put it another way: general relativity has gauge freedom, and we are choosing a convenient gauge.

With these choices, eqn. (\ref{mom3}) becomes
\begin{equation}
{{\tilde D}^i} \left ({{\tilde D} _i}{W_j}+{{\tilde D} _j}{W_i} - {\textstyle {\frac 2 3}}
{{\tilde \gamma} _{ij}} {{\tilde D}^k}{W_k} \right ) 
= - {{\tilde D}^i}{X_{ij}} \; - \; {{\tilde J}_j}
\label{mom4}
\end{equation}
This is a linear equation for $W_i$ that does not depend on $\psi$.  So the idea is to first solve eqn. (\ref{mom4}) for $W_i$ and then plug the result in to eqn. (\ref{ham3}) which is to be solved for $\psi$.  Eqn. (\ref{ham3}) is a somewhat complicated looking nonlinear equation.  But it is straightforward to solve it using standard numerical methods for nonlinear elliptic equations.  Therefore, for the rest of the paper we will only concentrate on how to solve eqn. (\ref{mom4}).

Equation (\ref{mom4}) is of the form operator acting on $W_i$ equals source, so the first thing we want to know is does the operator have a kernel?  That is, is there a vector $V_i$ for which
\begin{equation}
{{\tilde D}^i} \left ({{\tilde D} _i}{V_j}+{{\tilde D} _j}{V_i} - {\textstyle {\frac 2 3}}
{{\tilde \gamma} _{ij}} {{\tilde D}^k}{V_k} \right ) = 0
\label{momkernel}
\end{equation}
If there is no kernel, then the operator can be inverted and therefore there exists a unique solution of eqn. (\ref{mom4}). Multiplying eqn. (\ref{momkernel}) by $V^j$ and integrating over $\Sigma$ using ingegration by parts we have
\begin{equation}
{\int _\Sigma} \; \left ( {{\tilde D}^i} {V^j} \right )\left ({{\tilde D} _i}{V_j}+{{\tilde D} _j}{V_i} - {\textstyle {\frac 2 3}}
{{\tilde \gamma} _{ij}} {{\tilde D}^k}{V_k} \right ) = 0
\label{momkernelint}
\end{equation}
But this can only be the case if at each point we have
\begin{equation}
{{\tilde D} _i}{V_j}+{{\tilde D} _j}{V_i} - {\textstyle {\frac 2 3}}
{{\tilde \gamma} _{ij}} {{\tilde D}^k}{V_k}  = 0
\label{confKilling}
\end{equation}
Equation (\ref{confKilling}) is the conformal Killing equation.  Its solutions are conformal Killing vector fields.  But spaces with conformal Killing vectors are rare. Thus the conclusion for eqn. (\ref{mom4}) is that there is a general case (no conformal Killing vectors) in which there exists a unique solution, and then there is an exceptional case in which there is a conformal Killing vector.

\section{Cosmological Case}

Unfortunately, the exceptional case, although in some sense rare, is also the one of most relevance for cosmology.  Cosmological scalar perturbations have a conformally flat spatial metric.  A conformally flat metric has conformal Killing vector fields.  We are therefore led to investigate the exceptional case, and in fact to further specialize to the case where the conformally related metric ${\tilde \gamma}_{ij}$ is the flat metric $\delta _{ij}$ (i.e. the Kronecker delta).  Equation (\ref{mom4}) then becomes 
\begin{equation}
{\partial^i} \left ({\partial _i}{W_j}+{\partial _j}{W_i} - {\textstyle {\frac 2 3}}
{\delta _{ij}} {\partial^k}{W_k} \right ) 
= - {\partial^i}{X_{ij}} \; - \; {{\tilde J}_j}
\label{mom5}
\end{equation}
Here $\partial _i$ is the usual Cartesian coordinate derivative operator.

For linear equations where there is a kernel, we have the Fredholm alternative: any vector is expressed as the sum of two pieces, one in the kernel and one in the space orthogonal to the kernel (called the adjoint).  If the source is not in the adjoint, then the linear equation has no solutions.  If the source is in the adjoint, then the linear equation has multiple solutions, where any two solutions differ by something in the kernel.   

Our task in solving eqn. (\ref{mom5}) is therefore to first put conditions on the matter field initial data that insure that the right hand side of the equation is in the adjoint.  We must then find what is essentially the inverse of the operator on the adjoint space, in order to find a solution of the eqn. (\ref{mom5}).  There will be multiple solutions.  However, using the fact that any two solutions differ by something in the kernel, an examination of eqn. (\ref{Atilde}) shows that the two solutions give rise to the same ${\tilde A}_{ij}$, so in fact we can pick any solution, and it doesn't matter which one we pick.

A single mode in cosmological perturbation theory has spatial dependence only in the direction of propagation.
So we now further specialize to the case where there is dependence on only the $x$ coordinate.  We want initial data for a simulation with periodic boundary conditions, so we choose $x$ to be a periodic coordinate with period $2\pi$. We choose 
${W_y}={W_z}=0$  (That is we consider only choices of $X_{ij}$ for which the solution of eqn. (\ref{mom5}) gives ${W_y}={W_z}=0$).  Equation (\ref{mom5}) then becomes
\begin{equation}
{\textstyle {\frac 4 3}} {\frac {{d^2}{W_x}} {d{x^2}}} = - {\frac {d {X_{xx}}} {dx}} - {{\tilde J}_x}
\label{mom6}
\end{equation}

In some cases, the right hand side of eqn. (\ref{mom6}) is sufficiently simple that the equation can be solved in closed form.  However, other cases require a numerical method.  For similar equations, but ones without a kernel, the standard numerical method is to write the finite difference approximation of the equation as a matrix equation and then to perform an LU decomposition of the matrix.\cite{numericalrecipes}
However, eqn. (\ref{mom6}) does have a kernel, since a constant $W_x$ gives zero for the left hand side of the equation.  And indeed, application of the formula of \cite{numericalrecipes} to this case results in division by zero.  Instead we use a different type of LU decomposition method, described in the appendix, for the numerical solution of eqn. (\ref{mom6}).

Whether solved analytically or numerically, a solution of eqn. (\ref{mom6}) for $W_x$ gives rise to an expression for ${\tilde A}_{ij}$, which can in turn be used to solve eqn. (\ref{ham3}) for $\psi$.
The expression is ${{\tilde A}_{ij}}={X_{ij}}$ for $i \ne j$ and 
\begin{eqnarray}
{{\tilde A}_{xx}} = {X_{xx}} + {\textstyle {\frac 4 3}} {\frac {d{W_x}} {dx}}
\\
{{\tilde A}_{yy}} = {X_{yy}} - {\textstyle {\frac 2 3}} {\frac {d{W_x}} {dx}}
\\
{{\tilde A}_{zz}} = {X_{zz}} - {\textstyle {\frac 2 3}} {\frac {d{W_x}} {dx}}
\end{eqnarray}

\section{standard one-field case}

We now treat the case of cosmology with scalar field matter.  Here we will find that eqn. (\ref{mom6}) can be solved in closed form.  We want to find initial data that are as close as possible to a single mode of a cosmological scalar perturbation.  The stress-energy of the scalar field $\phi$ with potential $V(\phi)$ is
\begin{equation}
{T_{\mu \nu}} = {\nabla _\mu} \phi {\nabla _\nu} \phi \; - \; {g_{\mu \nu}} \, ( {\textstyle {\frac 1 2}}
{\nabla ^\alpha}\phi {\nabla _\alpha} \phi + V)
\label{stressenergy1}
\end{equation} 
Now using eqn. (\ref{stressenergy1}) in eqn. (\ref{current}) we find
\begin{equation}
{{\tilde J}_j} = {\psi ^6} P {\partial _j}\phi
\label{current2}
\end{equation}
where the quantity $P$ is defined by $P={n^\mu}{\nabla _\mu}\phi$.  To make ${\tilde J}_j$ independent of $\psi$ we define the quantity $Q$ by
\begin{equation}
Q = {\psi ^6} P
\label{Qdef}
\end{equation}
which leads to 
\begin{equation}
{{\tilde J}_j} = Q {\partial _j}\phi
\label{current3}
\end{equation}
So we specify $Q$ and it is only at the end, when we have numerically solved for $\psi$ that we know the stress-energy.

We will find the initial values for $Q$ and $\phi$ of a cosmological scalar perturbation, and use those in eqns. (\ref{mom6}) and (\ref{current3}) to find the general relativity initial data.  

The background Friedmann-Lemaitre-Robertson-Walker (FLRW) spacetime has the line element
\begin{equation}
d {s^2} = - d {t^2} \; + \; {a^2}(t) (d {x^2} + d {y^2} + d {z^2} ) 
\label{FLRW}
\end{equation}
We will denote quantities in the background with a subscript zero, and use an overdot for derivative with respect to $t$.  The Hubble parameter $H$ is given by $H={\dot a}/a$.  Then we have
\begin{eqnarray}
{K_0} = - 3 H
\\
{Q_0} = {a^3} {{\dot \phi}_0}
\end{eqnarray}

A single mode of the scalar field is usually written as a function of time multiplied by $e^{iqx}$, with the notion that since the equations are linear, we can do all our computations with the complex mode and at the end of the day we will take the real part.  However, ${\tilde J}_j$ is quadratic in the scalar field, not linear, so we will write our modes as real quantities from the start.  Since we have in mind initial data for simulations with periodic boundary conditions, we will choose $x$ to be a periodic variable going from $0$ to $2\pi$.  Therefore $q$ will be an integer.  The quantities $\phi$ and $Q$ take the form
\begin{eqnarray}
\phi = {\phi_0} \; + \; {c_1} \cos(q x) \; + \; {c_2} \sin (q x)
\label{onefieldphi}
\\
Q = {Q_0}  \; + \; {c_3} \cos(q x) \; + \; {c_4} \sin (q x) 
\label{onefieldQ}
\end{eqnarray}
Where ${c_1}, \, {c_2}, \, {c_3}$ and $c_4$ are constants

Cosmological scalar perturbations have ${X_{ij}}=0$, so eqn. (\ref{mom6}) becomes
\begin{equation}
{\textstyle {\frac 4 3}} {\frac {{d^2}{W_x}} {d{x^2}}} = - {{\tilde J}_x}
\label{mom7}
\end{equation}
Using eqns. (\ref{onefieldphi}-\ref{onefieldQ}) in eqn. (\ref{current3}) we obtain
\begin{eqnarray}
- {{\tilde J}_x} &=&  q ( {Q_0} \; + \; {c_3} \cos(q x) \; + \; {c_4} \sin (q x) ) (
{c_1} \sin(q x) \; - \; {c_2} \cos (q x))
\\
&=& {Q_0} q \left [ {c_1} \sin(q x) \; - \; {c_2} \cos (q x) \right ]
\nonumber
\\
 &+& {\textstyle {\frac 1 2}}  q \left [ 
({c_1}{c_4}-{c_2}{c_3}) - ({c_1}{c_4}+{c_2}{c_3}) \cos (2 q x) + ({c_1}{c_3}-{c_2}{c_4}) \sin (2 q x) \right ] 
\label{current4}
\end{eqnarray}
The requirement that the source be in the adjoint, means that the constant term on the right hand side of eqn. (\ref{current4}) must vanish.  That is, we must require
\begin{equation}
{c_1}{c_4}={c_2}{c_3}
\end{equation}
This sort of constraint on the freedom to specify a cosmological perturbation is known as an integral constraint.\cite{Traschen}

Using eqn. (\ref{current4}) in eqn. (\ref{mom7}) and integrating, we obtain
\begin{eqnarray}
{\textstyle {\frac 4 3}} {\frac {d{W_x}} {dx}} &=&  - \, {Q_0}  \left [ {c_1} \cos(q x) \; + \; {c_2} \sin (q x) \right ]
\nonumber
\\
&-& \;  {\textstyle {\frac 1 4}}  \left [ 
 ({c_1}{c_4}+{c_2}{c_3}) \sin (2 q x) + ({c_1}{c_3}-{c_2}{c_4}) \cos (2 q x) \right ] 
\end{eqnarray}
This is our solution of the momentum constraint equation.

We will now express the parameters $({c_1},{c_2},{c_3},{c_4})$ in terms of the standard cosmological perturbation theory\cite{BST,weinberg} in Newtonian gauge.  

The line element in Newtonian gauge takes the form  
\begin{equation}
d {s^2} = - (1 + 2\Psi) d {t^2}  \; + \; {a^2} (1 - 2 \Psi){\delta _{ij}} \, d{x^i} \, d{x^j}
\label{metric2}
\end{equation}
Where $\Psi$ is the cosmological Newtonian potential.

The scalar field in Newtonian gauge takes the form
\begin{equation}
{\phi_N} = {\phi_0} + \alpha (t) \cos (q x) + \beta (t) \sin (q x)
\label{Newtonianphi}
\end{equation}

From eqn. (\ref{metric2}) we find that $Q$ and $K$ in Newtonian gauge are
\begin{eqnarray}
{Q_N} &=& {a^3}{{\dot \phi}_0} (1 - 4 \Psi) + {a^3} ( {\dot \alpha} \cos (q x) + {\dot \beta}\sin (q x))
\label{NewtonianQ}
\\
{K_N} &=& - 3 H + 3 ( {\dot \Psi} + H \Psi )
\label{NewtonianK}
\end{eqnarray}
It is clear from eqn. (\ref{NewtonianK}) that $K_N$ has dependence on the spatial coordinates, and therefore that Newtonian gauge is not CMC gauge.  However, we can transform to CMC gauge through the use of a gauge transformation.  In general relativistic perturbation theory, for every vector field $\xi^\mu$ there is a gauge transformation that consists of adding to each quantity Lie derivative with respect to $\xi ^\mu$ of the background quantity.  We will choose our vector field to have only a time component.  The gauge transformed $K$ is then
\begin{equation}
K = {K_N} + {{\cal L}_\xi} {K_0} = - 3 H + 3 ( {\dot \Psi} + H \Psi ) + {\xi^t} {\partial _t} (- 3 H)
= 3( - H + {\dot \Psi} + H \Psi - {\xi ^t} {\dot H} )
\end{equation}
Thus to make $K$ spatially constant, we choose $\xi ^t$ to be
\begin{equation}
{\xi ^t} = {\frac {{\dot \Psi} + H \Psi} {\dot H}}
\end{equation}
However a standard result of cosmological perturbation theory in Newtonian gauge is\cite{weinberg}
\begin{equation}
{\dot \Psi} + H \Psi = {\textstyle {\frac 1 2}} {{\dot \phi}_0} ({\phi_N} - {\phi_0})
\end{equation}
so we find
\begin{equation}
{\xi ^t} = {\frac {{\dot \phi}_0} {2 {\dot H}}} ({\phi_N} - {\phi_0})
\end{equation}
Applying the gauge transformaton, we find that the scalar field in CMC gauge is
\begin{eqnarray}
\phi &=& {\phi _N} + {{\cal L}_\xi} {\phi _0} = {\phi _N} + {\xi ^t} {{\dot \phi}_0}
= {\phi _0} + \left ( 1 +  {\frac {{\dot \phi}_0 ^2} {2 {\dot H}}} \right ) ({\phi_N} - {\phi_0})
\nonumber 
\\
&=&  {\phi _0} + \left ( 1 +  {\frac {{\dot \phi}_0 ^2} {2 {\dot H}}} \right ) (\alpha \cos (qx) + \beta \sin (qx))
\label{CMCphi}
\end{eqnarray}
Comparing eqns. (\ref{onefieldphi}) and (\ref{CMCphi}) we see that two of the parameters of our momentum constraint solution are given by 
\begin{equation}
{c_1} =  \left ( 1 +  {\frac {{\dot \phi}_0 ^2} {2 {\dot H}}} \right ) \alpha , \; \; \; \; 
{c_2} =  \left ( 1 +  {\frac {{\dot \phi}_0 ^2} {2 {\dot H}}} \right ) \beta 
\label{c12solution}
\end{equation}
where all quantities are evaluated at the time $t_0$ of our initial data.

We now find the quantity $Q$ in CMC gauge.  We have 
\begin{equation}
{Q} = {Q_N} + {{\cal L}_\xi} {Q_0} = {Q_N} - {Q_0} {\frac {{V'}({\phi _0})} {{\dot \phi}_0}} {\xi ^t}
= {Q_N} - {Q_0} {\frac {{V'}({\phi _0})} {2 {\dot H}}} (\phi - {\phi_N})
\label{CMCQ}
\end{equation}
where we have used the equation of motion for the background scalar field.

To evaluate the term proportional to $\Psi$ in the expression of eqn. (\ref{onefieldQ}) for $Q_N$, we use the following result of cosmological perturbation theory in Newtonian gauge:\cite{weinberg}
\begin{equation}
({\dot H} + {q^2}/{a^2}) \Psi = {\textstyle {\frac 1 2}} {{\ddot \phi}_0}({\phi _N} - {\phi _0}) \; - \;
{\textstyle {\frac 1 2}} {{\dot \phi}_0}({{\dot \phi}_N} - {{\dot \phi}_0}) 
\label{weinbergPsi}
\end{equation}
Combining eqns. (\ref{onefieldQ}), (\ref{CMCQ}) and (\ref{weinbergPsi}) we obtain
\begin{eqnarray}
Q = {Q_0} \; &+& \; {a^3} \left ( 1 \; + \; {\frac {2 {{\dot \phi}_0 ^2}}
{{\dot H} + {q^2}/{a^2}}} \right ) ({\dot \alpha} \cos (qx) + {\dot \beta} \sin (qx))
\nonumber
\\
&-& \; {a^3}  {{\dot \phi}_0} \left ( {\frac {{V'}({\phi_0})} {2 {\dot H}}} \; + \; {\frac {2 {{\ddot \phi}_0}} {{\dot H} + {q^2}/{a^2}}} \right ) ( \alpha \cos (qx) + \beta \sin (qx)) 
\label{CMCQ2}
\end{eqnarray}
Comparing eqns. (\ref{onefieldQ}) and (\ref{CMCQ2}) we find that the remaining two parameters of our momentum constraint solution are given by
\begin{eqnarray}
{c_3} = {a^3} \left ( 1 \; + \; {\frac {2 {{\dot \phi}_0 ^2}}
{{\dot H} + {q^2}/{a^2}}} \right ) {\dot \alpha} 
\; - \; {a^3} {{\dot \phi}_0} \left ( {\frac {{V'}({\phi_0})} {2 {\dot H}}} \; + \; {\frac {2 {{\ddot \phi}_0}} {{\dot H} + {q^2}/{a^2}}} \right ) \alpha
\nonumber
\\
{c_4} = {a^3} \left ( 1  \; + \; {\frac {2 {{\dot \phi}_0 ^2}}
{{\dot H} + {q^2}/{a^2}}} \right ) {\dot \beta} 
\; - \; {a^3} {{\dot \phi}_0} \left ( {\frac {{V'}({\phi_0})} {2 {\dot H}}} \; + \; {\frac {2 {{\ddot \phi}_0}} {{\dot H} + {q^2}/{a^2}}} \right ) \beta
\label{c34solution}
\end{eqnarray}
where all quantities are evaluated at the time $t_0$ of our initial data.  

Using eqns. (\ref{c12solution}) and (\ref{c34solution}), we see that the constraint on the parameters ${c_1}{c_4}={c_2}{c_3}$ becomes
\begin{equation}
\alpha {\dot \beta} = \beta {\dot \alpha}
\end{equation}

\section{Ekpyrotic two-field case}

We now treat the case of the ekpyrotic two-field model.\cite{twofield} In this model there is a scalar field $\phi$ with a potential $V(\phi)$ and thus the same stress-energy as in eqn. (\ref{stressenergy1}).  However, there is also a second scalar field $\chi$ whose kinetic term is coupled to the first scalar field through a function $\kappa (\phi)$.  In the ekpyrotic scenario, $\phi$ causes the smoothing during a contracting phase prior to the bounce into the Big Bang, while $\phi$ and $\chi$ together insure the appropriate spectrum of perturbations.  The combined stress-energy of the two fields is 
\begin{eqnarray}
{T_{\mu \nu}} = {\nabla _\mu} \phi {\nabla _\nu} \phi \; - \; {g_{\mu \nu}} \, ( {\textstyle {\frac 1 2}}
{\nabla ^\alpha}\phi {\nabla _\alpha} \phi + V)
\nonumber
\\
+ \; \kappa (\phi) \left [ {\nabla _\mu} \chi {\nabla _\nu} \chi \; - \; {\textstyle {\frac 1 2}} \, {g_{\mu \nu}} \, {\nabla ^\alpha}\chi {\nabla _\alpha} \chi  \right ] 
\label{stressenergy2}
\end{eqnarray} 

\begin{figure}
\centering
\includegraphics[width=4.5in]{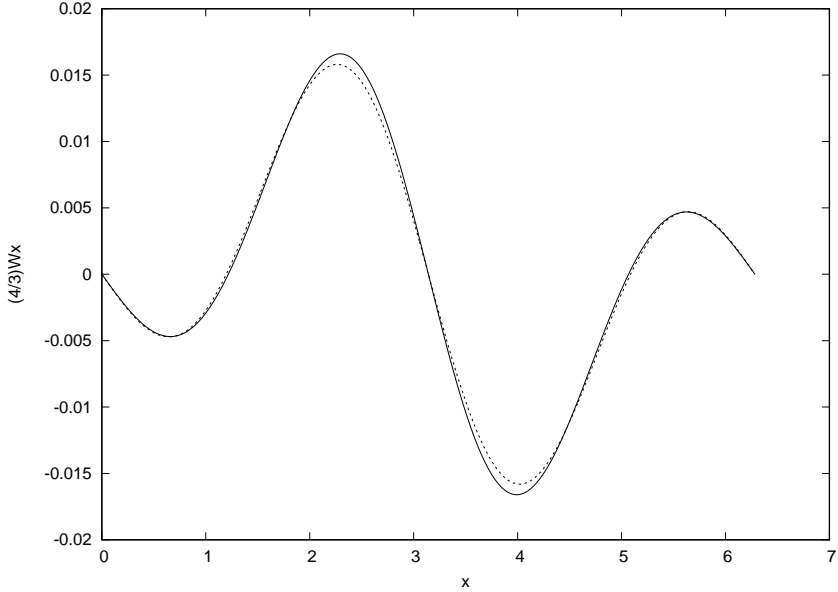}
\caption{$(4/3){W_x}$ vs. $x$ for the numerical method (solid line) and perturbative method (dashed line) for weak initial data}
\label{weak}
\end{figure}

\begin{figure}
\centering
\includegraphics[width=4.5in]{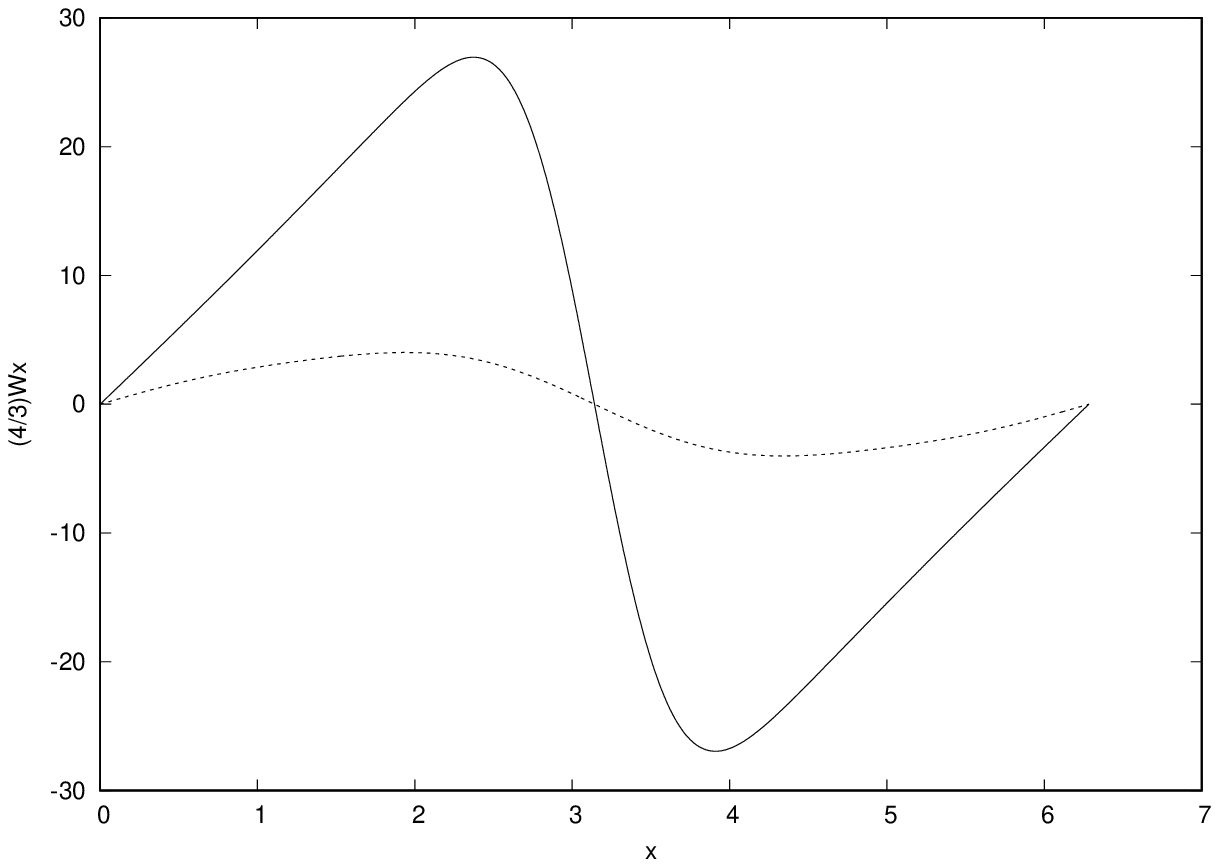}
\caption{$(4/3){W_x}$ vs. $x$ for the numerical method (solid line) and perturbative method (dashed line) for strong initial data}
\label{strong}
\end{figure}

As before, we define $P$ and $Q$ by $P={n^\mu}{\nabla _\mu}\phi$ and $Q = {\psi ^6} P$.  However, we also define $P_\chi$ and $Q_\chi$ by ${P_\chi}={n^\mu}{\nabla _\mu}\chi$ and ${Q_\chi} = {\psi ^6}{P_\chi}$.  
Since we are concerned with scalar modes, we will choose ${X_{ij}}=0$.  
Then the momentum constraint once again takes the form
\begin{equation}
{\textstyle {\frac 4 3}} {\frac {{d^2}{W_x}} {d{x^2}}} = - {{\tilde J}_x}
\label{mom8}
\end{equation}
But now with ${\tilde J}_x$ taking the form
\begin{equation}
{{\tilde J}_x} = Q {\partial _x} \phi + \kappa (\phi) {Q_\chi} {\partial _x} \chi
\end{equation}

In this case, we are not so much concerned with matching a particular perturbative mode, but rather with coming up with a class of initial data, not necessarily small, of sufficient generality to allow a thorough numerical exploration of the two-field ekpyrotic scenario.  The condition needed for a solution of eqn. (\ref{mom8}), namely that ${\tilde J}_x$ be in the adjoint, becomes
\begin{equation}
{\int _0 ^{2\pi}} \; dx \; {{\tilde J}_x} = 0
\label{condition}
\end{equation}
One simple way to satisfy this condition is to make $\phi, \, \chi, \, Q$ and $Q_\chi$ functions of $\cos x$.  In this way, both $Q{\partial _x}\phi$ and $\kappa (\phi) {Q_\chi}{\partial _x}\chi$ become odd functions of $x$, whose integral over one period therefore vanishes.  We will take the usual choice for 
$\kappa (\phi)$ of
\begin{equation}
\kappa (\phi) = {e^{- c \phi}}
\end{equation}
where $c$ is a constant.  For non-perturbative initial data, we cannot solve eqn. (\ref{mom8}) in closed form.  Therefore instead we use the numerical method presented in the appendix.  If we were doing a perturbative treatment, we would replace $e^{-c\phi}$ with $1-c\phi$ and solve eqn. (\ref{mom8}) using the analytic methods of the previous section.  Figures (\ref{weak}) and (\ref{strong}) shows the results of such a numerical solution.  
Here we have used $\phi, \, \chi, \, Q$ and $Q_\chi$ of the form: $\phi = {c_0} \cos (q x), \; Q = {c_1} \cos (q x), \; \chi = {d_0} \cos (q x), \; {Q_\chi} = {d_1} \cos (q x) $.  We plot the results of the numerical treatment in a solid line and the results of the corresponding perturbative-analytic treatment in a dashed line.  In figure (\ref{weak}) we pick parameters $c=5, \, q=1, \, {c_0}=0.1, \, {c_1} = 0.2, \, {d_0} = 0.2, \, {d_1} = 0.3$ which correspond to weak initial data.  Note that in this case the perturbative result is quite close to the numerical result.  In contrast, in figure (\ref{strong}) we pick parameters
$c=5, \, q=1, \, {c_0}=1.0, \, {c_1} = 1.4, \, {d_0} = 2.0, \, {d_1} = 1.6$ corresponding to much stronger initial data.  Here the perturbative result is not at all a good approximation for the full numerical treatment, and so the numerical method is definitely needed.

\section{Conclusion}

We have provided methods to generate more extensive sets of initial data for numerical relativity simulations of inhomogeneous cosmologies.  The sort of data needed for inhomogeneous cosmologies constitute an exceptional case within the York method for finding general relativity initial data.  Because it is exceptional, this case cannot be treated using the standard numerical methods.  Nonetheless, we have found some situations where the problem can be solved in closed form.  And for the situations that cannot be treated in closed form, we have found a numerical method, a subtle modification of the standard LU decomposition method, that works.  

Typically the goal of numerical relativity simulations of inhomogeneous cosmologies is to make assertions about what outcomes result from ``generic'' initial conditions.  But this means that the wider the class of initial data used for the simulations, the more confidently one can assert that the simulations give the generic outcome.  It would be interesting to repeat some of the simulations of inhomogeneous cosmologies (e.g. some of the ones given in the references of this paper) with our more general initial data to see if the conclusions about outcomes remain the same.

\section*{Acknowledgments}

We would like to thank Anna Ijjas, Paul Steinhardt, and Frans Pretorius for helpful discussions.
David Garfinkle thanks Princeton University for hospitality, and acknowledges support from NSF Grant PHY-1806219. 

\appendix

\section{numerical method}

We need to numerically solve an equation of the form
\begin{equation}
{\frac {{d^2}f} {d{x^2}}} = g
\label{laplace1}
\end{equation}
on a grid with periodic boundary conditions.  We pick $N$ grid points with spacing $\Delta$ and denote with a subscipt $i$ the value of the function at grid point $i$.  Using centered differences, eqn. (\ref{laplace1}) becomes
\begin{equation}
{\frac {{f_{i+1}}+{f_{i-1}}-2{f_i}} {\Delta ^2}} = {g_i}
\label{laplace2}
\end{equation}
This equation can be used at all grid points except grid points $1$ and $N$.  To evaluate eqn. (\ref{laplace1}) at these points, we add two ghost zones, grid points $0$ and $N+1$ that implement the periodic boundary conditions: ${f_0}={f_N}$ and ${f_{N+1}}={f_1}$. We then find
\begin{eqnarray}
{\Delta ^2} {g_1} = {f_2} + {f_0} - 2 {f_1} = {f_2} + {f_N} - 2 {f_1}
\\
{\Delta ^2} {g_N} = {f_{N+1}} + {f_{N-1}} - 2 {f_N} = {f_1} + {f_{N-1}} - 2 {f_N}
\end{eqnarray}

Using the notation $\left | f \right >$ for the column vector of $f_i$ and similarly for 
$\left | g \right >$ we find that eqn. (\ref{laplace2}) with periodic boundary conditions applied can be written as the matrix equation $ A \left | f \right >  = {\Delta ^2} \left | g \right > $ where for definiteness we display the matrix $A$ for the case $N=4$.
\begin{equation}
A = 
\begin{pmatrix}
-2 & 1 & 0 & 1 \\
1 & -2 & 1 & 0 \\
0 & 1 & -2 & 1 \\
1 & 0 & 1 & -2 
\end{pmatrix}
\label{Amatrix}
\end{equation}

If $A$ were invertible, we could solve for $ \left | f \right > $ by multiplying both sides of the equation 
$ A \left | f \right >  = {\Delta ^2} \left | g \right > $ by $A^{-1}$.  However, it is easy to see that $A$ is not invertible, since it annihilates the vector $\left | f \right > $ where all the $f_i$ are equal to the same constant.  This is just the finite difference version of the statement that the operator 
${d^2}/d{x^2}$ annihilates the function $f$ that is a constant. 

For an invertible matrix, there is a standard decomposition of the matrix into lower and upper triangular matricies (called LU decomposition) that allows a convenient algorithm\cite{numericalrecipes} for solving the system of linear equations associated with the matrix.  The matrix $A$ is not invertible, but nonetheless, we have an analog of the LU decomposition, which we display for the $N=4$ case: $A=LU$ where
\begin{equation}
L = 
\begin{pmatrix}
-1 & 0 & 0 & 1 \\
1 & -1 & 0 & 0 \\
0 & 1 & -1 & 0 \\
0 & 0 & 1 & -1 
\end{pmatrix}
\label{Lmatrix}
\end{equation} 
\begin{equation}
U = 
\begin{pmatrix}
1 & -1 & 0 & 0 \\
0 & 1 & -1 & 0 \\
0 & 0 & 1 & -1 \\
-1 & 0 & 0 & 1 
\end{pmatrix}
\label{Umatrix}
\end{equation}
Note that despite their names, the matrix $L$ is not lower triangular, because of the entry in the upper right hand corner, and the matrix $U$ is not upper triangular because of the entry in the lower left hand corner.

As with standard LU decomposition, the idea is that to solve the equation 
$LU\left | x \right > = \left | r \right > $ for $\left | x \right > $, we first solve
$L \left | y \right > = \left | r \right > $ for $\left | y \right > $ and then solve
$U \left | x \right > = \left | y \right > $ for $\left | x \right > $.  We will work out this problem explicitly for the $N=4$ case illustrated in eqns. (\ref{Amatrix}-\ref{Umatrix}).  Then we will describe the corresponding algorithm for general $N$.  The equation $L \left | y \right > = \left | r \right > $ becomes the following set of linear equations:
\begin{eqnarray}
-{y_1} + {y_4} &=& {r_1}
\label{L1}
\\
{y_1}-{y_2} &=& {r_2}
\label{L2}
\\
{y_2}-{y_3} &=& {r_3}
\label{L3}
\\
{y_3}-{y_4} &=& {r_4}
\label{L4}
\end{eqnarray}
Adding eqns. (\ref{L1}-\ref{L4}) we obtain ${r_1}+{r_2}+{r_3}+{r_4}=0$.  In other words $\left | r \right > $
must be in the adjoint, which is what the Fredholm alternative tells us needs to be true anyway if there is to be a solution to the original problem $ A\left | x \right > = \left | r \right > $. 

Notice that the left hand sides of eqns. (\ref{L1}-\ref{L4}) are each differences of two $y_i$.  This means that if we have a solution of these equations, then we can obtain another solution simply by adding the same constant to each $y_i$.  We will exploit this freedom to choose ${y_4}=0$.  Note that eqn. (\ref{L1}) then yields ${y_1}=-{r_1}$.  But knowing $y_1$ now allows us to solve eqn. (\ref{L2}) for $y_2$, which in turn allows us to solve eqn. (\ref{L3}) for $y_3$.  This solution for the $y_i$ is then
\begin{equation}
\left | y \right > = 
\begin{pmatrix}
- {r_1} \\
- ({r_1}+{r_2}) \\
- ({r_1}+{r_2}+{r_3}) \\
0 
\end{pmatrix}
\label{yprelim}
\end{equation}
Note that the average value of the $y_i$ is then ${\bar y} = (-1/4)(3{r_1}+2{r_2}+{r_3})$.  We will produce a new solution by subtracting this average from each $y_i$ and thus have a solution where the sum of the $y_i$ vanishes.  (as we will soon see, we will need this solution in order to solve the equation 
$U \left | x \right > = \left | y \right > $).  The new solution is
\begin{equation}
\left | y \right > = {\frac 1 4}
\begin{pmatrix}
- {r_1} + 2 {r_2} + {r_3}\\
- {r_1}-2{r_2} +{r_3} \\
- {r_1}-2{r_2}-3{r_3} \\
3{r_1}+2{r_2}+{r_3} 
\end{pmatrix}
\label{ysoln}
\end{equation}

The equation $U \left | x \right > = \left | y \right > $ becomes the following set of linear equations:
\begin{eqnarray}
{x_1} - {x_2} &=& {y_1}
\label{U1}
\\
{x_2}-{x_3} &=& {y_2}
\label{U2}
\\
{x_3}-{x_4} &=& {y_3}
\label{U3}
\\
-{x_1}+{x_4} &=& {y_4}
\label{U4}
\end{eqnarray}
Adding eqns. (\ref{U1}-\ref{U4}) we obtain ${y_1}+{y_2}+{y_3}+{y_4}=0$.  In other words we did need to impose the condition that $\left | y \right > $
is in the adjoint on the previous solution.

Since the left hand sides of eqns. (\ref{U1}-\ref{U4}) are each differences of two $x_i$, we can obtain from any solution another solution simply by adding the same constant to each $x_i$.  We will exploit this freedom to choose ${x_1}=0$.  Note that eqn. (\ref{U4}) then yields ${x_4}={y_4}$.  But knowing $x_4$ now allows us to solve eqn. (\ref{U3}) for $x_3$, which in turn allows us to solve eqn. (\ref{U2}) for $x_2$.  This solution for the $x_i$ is then
\begin{equation}
\left | x \right > = 
\begin{pmatrix}
0 \\
{y_2}+{y_3}+{y_4} \\
{y_3}+{y_4} \\
{y_4} 
\end{pmatrix}
\label{xprelim}
\end{equation}
Note that the average value of the $x_i$ is then ${\bar x} = (1/4)({y_2}+2{y_3}+3{y_4})$.  Though not strictly necessary, we will procede in analogy to our previous method for finding $\left | y \right >$ and produce a new solution for $\left | x \right >$ by subtracting this average from each $x_i$ and thus have a solution where the sum of the $x_i$ vanishes.  The new solution is
\begin{equation}
\left | x \right > = {\frac 1 4}
\begin{pmatrix}
-{y_2}-2{y_3}-3{y_4}\\
3{y_2}+2{y_3}+{y_4} \\
-{y_2}+2{y_3}+{y_4} \\
-{y_2}-2{y_3}+{y_4} 
\end{pmatrix}
\label{xsoln1}
\end{equation}

Finally, using eqn. (\ref{ysoln}) in eqn. (\ref{xsoln1}) we obtain the solution to the original problem 
$LU \left | x \right > = \left | r \right >$.
\begin{equation}
\left | x \right > = {\frac 1 8}
\begin{pmatrix}
-3{r_1}+{r_3}\\
-{r_1}-4{r_2}-{r_3} \\
{r_1}-3{r_3} \\
3{r_1}+4{r_2}+3{r_3} 
\end{pmatrix}
\label{xsoln2}
\end{equation}
This solution can also be expressed in a slightly more natural looking way using ${r_1}+{r_2}+{r_3}+{r_4}=0$ as
\begin{equation}
\left | x \right > = {\frac 1 8}
\begin{pmatrix}
{r_3}-3{r_1}\\
{r_4}-3{r_2} \\
{r_1}-3{r_3} \\
{r_2}-3{r_4} 
\end{pmatrix}
\label{xsoln3}
\end{equation}

We now describe the general form of the algorithm to obtain this solution (i.e. for general $N$, not restricted to $N=4$).  The kernel of $A$ consists of all $\left | f \right > $ where the $f_i$ all have the same values.  The adjoint of $A$ consists of all $\left | f \right > $ where ${\sum _{i=1} ^N} {f_i} = 0$.  This kernel of $A$ is also the kernel of $L$ and $U$, and the adjoint of $A$ is also the adjoint of $L$ and $U$.  The vector $\left | r \right > $, must be in the adjoint, or there is no solution of $L \left | y \right > = \left | r \right > $.  But if $\left | r \right > $ is in the adjoint, then there are multiple solutions for $\left | y \right >$ each differing by something in the kernel.  We make use of this freedom to choose ${y_N}=0$.  It then follows that ${y_1}=-{r_1}$; and that ${y_{i+1}} = {y_i} - {r_{i+1}}$, which we iteratively solve in succession for ${y_2},{y_3},\dots {y_{N-1}}$.  This $\left | y \right >$ is generally not in the adjoint, which would make it impossible to solve
$U \left | x \right > = \left | y \right > $.  However, we turn it into a solution in the adjoint by subtracting the appropriate vector in the kernel.  That is, we find the average $\bar y$ of the $y_i$ and then subtract $\bar y$ from each $y_i$ to make our new vector $\left | y \right >$.  Now we use the same sort of procedure to solve 
$U \left | x \right > = \left | y \right > $.  We use the freedom to add something in the kernel to choose 
${x_1}=0$.  We then have ${x_N}={y_N}$, as well as ${x_{i-1}}={x_i}+{y_{i-1}}$ which we solve iteratively for ${x_{N-1}},{x_{N-2}},\dots {x_2}$.  This $\left | x \right >$ is a solution of the equation
$A\left | x \right > = \left | r \right >$ but we go ahead and produce a solution in the adjoint by subtracting $\bar x$ from each $x_i$.

This algorithm may sound a bit complicated, but it is straightforward to program and the resulting code is about the same length as the general description given above of the algorithm.

\end{document}